\documentclass[runningheads]{llncs}
\usepackage{graphicx}
\usepackage{lscape}
\usepackage{color}
\usepackage[table]{xcolor}
\usepackage{colortbl}
\usepackage[utf8]{inputenc}
\usepackage{csquotes}
\usepackage{listings}
\usepackage[capitalise,nameinlink]{cleveref}
\usepackage[caption=false]{subfig}
\usepackage{tabularx,booktabs}
\usepackage{multirow}
\usepackage{bm}
\usepackage{todonotes}
\usepackage{slashbox}
\usepackage[graphicx]{realboxes}

\newcommand{\images}{images/}

\newcommand{\eg}{e.\,g.,\ }
\newcommand{\ie}{i.\,e.,\ }
\newcommand{\etal}{et\,al.}
\mathchardef\mhyphen="2D

\newcommand{\schemnew}{SchemEX+U+I}
\newcommand{\schemnewLong}{SchemEX with \textbf{U}nions of sameAs instances and RDFS \textbf{I}nferencing}
\newcommand{\schemnewo}{SchemEX+U+oI}
\newcommand{\schemnewp}{SchemEX+U+pI}

\newcommand{\hl}[1]{\textcolor{black}{#1}}
\newcommand{\rotateHeading}[1]{\rotatebox[origin=b]{90}{\textbf{#1}}}

\usepackage{eso-pic}

\definecolor{Gray}{gray}{0.925} % Table
\definecolor{Preprint}{rgb}{.63,.79,.95}

\newcommand{\preprintBanner}{
\AddToShipoutPictureFG*{\put(\LenToUnit{0.5\paperwidth},\LenToUnit{0.925\paperheight}){\makebox[0pt][c]{
\renewcommand{\arraystretch}{1.5}
\setlength{\tabcolsep}{18pt}
\rowcolors{1}{Preprint}{Gray}
\begin{tabular}{|p{0.55\paperwidth}|} \hline
\textbf{Extended Technical Report} \\ \hline
\footnotesize
T. Blume, A. Scherp, ``Indexing Data on the Web: A Comparison of
Schema-level Indices for Data Search'' in \textit{Proceedings of Database and Expert Systems Applications - 31th International Conference, {DEXA}}, Bratislava, Slovakia, September 14-17, 2020.\\ \hline
\end{tabular}}}}
}
\begin{document}
\renewcommand{\thelstlisting}{\arabic{lstlisting}}

\title{Indexing Data on the Web: A Comparison of Schema-level Indices for Data Search}
\subtitle{Extended Technical Report}
\titlerunning{Schema-level Indices for Data Search}
%
% \orcidID{0000-0001-6970-9489}
% \orcidID{0000-0002-2653-9245}
\author{Till Blume\inst{1}  \and
Ansgar Scherp\inst{2}}
\authorrunning{T. Blume and A. Scherp}
\institute{Kiel University, Germany\\
\email{tbl@informatik.uni-kiel.de}\\
 \and
Ulm University, Germany\\
\email{ansgar.scherp@uni-ulm.de}}
\maketitle              % typeset the header of the contribution

\preprintBanner

\begin{abstract}
Indexing the Web of Data offers many opportunities, in particular, to find and explore data sources. One major design decision when indexing the Web of Data is to find a suitable index model, i.e., how to index and summarize data. Various efforts have been conducted to develop specific index models for a given task.
With each index model designed, implemented, and evaluated independently, it remains difficult to judge whether an approach generalizes well to another task, set of queries, or dataset. 
In this work, we empirically evaluate six representative index models with unique feature combinations. Among them is a new index model incorporating inferencing over RDFS and \texttt{owl:sameAs}. 
We implement all index models for the first time into a single, stream-based framework. 
We evaluate variations of the index models considering sub-graphs of size $0$, $1$, and $2$ hops on two large, real-world datasets.
We evaluate the quality of the indices regarding the compression ratio, summarization ratio, and F1-score denoting the approximation quality of the stream-based index computation.
The experiments reveal huge variations in compression ratio, summarization ratio, and approximation quality for different index models, queries, and datasets.
However, we observe meaningful correlations in the results that help to determine the right index model for a given task, type of query, and dataset.
\end{abstract}

\section{Introduction}
Graph indices are well-established to efficiently manage large heterogeneous graphs like the Web of Data.
In general, one can distinguish instance-level indices and schema-level indices for the Web of Data.
Instance-level indices focus on finding specific data instances~\cite{Tran:2009:SSU:1615258.1615264,DBLP:conf/rweb/HoseSTW11,DBLP:conf/ekaw/LeiUM06}, \eg searching for a specific book by its title such as \enquote{Towards a clean air policy}.
In contrast, schema-level indices (short: \emph{SLI}) support structural queries, \eg searching for data instances with the property \textit{dct:creator} and RDF type \textit{bibo:book}~\cite{Lodatio:2013}.
An \emph{SLI model} defines how and which combinations of types and properties are indexed, \ie how data instances are summarized and which queries are supported by the index. 

In the past, various SLI models have been developed for different tasks such as data exploration~\cite{DBLP:conf/webi/BenedettiBP15,DBLP:conf/semweb/Mihindukulasooriya15,pietriga2018browsing,DBLP:conf/esws/SpahiuPPRM16a}, query size estimation~\cite{CharacteristicSets:2011}, vocabulary terms recommendation~\cite{DBLP:conf/esws/SchaibleGS16}, related entity retrieval~\cite{SemSets:2012}, data search~\cite{Lodatio:2013}, and others.
The task of data search is to find (sub-) graphs on the Web that match a given schema structure.
Given a structural query as shown in \cref{lst:example-query}, an SLI can return URIs of data sources $?ds$ with data instances containing bibliographic metadata.

\begin{lstlisting}[firstnumber=1, caption={Structural query to find datasources ?ds containing information about books that have a creator that is an agent. },label=lst:example-query,breaklines=true,
breakindent=5pt,basicstyle=\scriptsize]
SELECT ?ds WHERE { 
   ?ds rdf:type bibo:book ;
       dct:creator foaf:Agent .	}
\end{lstlisting}

\begin{figure*}[!h]
\centering
\includegraphics[width=0.99\linewidth]{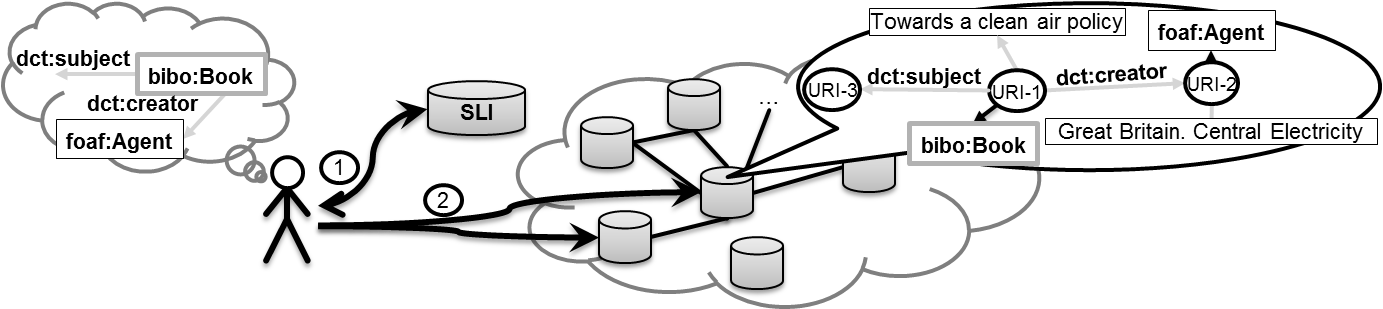}
\caption{\label{fig:SLI-setting}Finding data sources on the Web using a schema-level index (SLI).
A structural query is executed over an SLI to identify relevant data sources (1). 
Subsequently, the data sources are accessed to retrieve actual data instances (2).}
\end{figure*}

This use of SLIs for data search on the Web of Data is illustrated in \cref{fig:SLI-setting}.
In a first step, an SLI is queried to identify relevant data sources, which then in a second step are accessed via HTTP get requests to download the actual data instances.
Search systems like LODatio~\cite{Lodatio:2013}, LODeX~\cite{DBLP:conf/webi/BenedettiBP15}, Loupe~\cite{DBLP:conf/semweb/Mihindukulasooriya15}, and LODatlas~\cite{pietriga2018browsing} rely on SLI to offer a search for relevant data sources or exploration of data sources. 

The problem is that all SLI models were designed, implemented, and evaluated for their individual task only, using different queries, datasets, and metrics.
Our hypothesis is that there is no SLI model that fits all tasks and that the performance of the specific SLI depends on the specific types of queries and characteristics of the datasets.
However, so far only very limited work has been done on understanding the behavior of SLI models in different contexts.
With each SLI model evaluated independently in a specific context, it remains difficult to judge whether an approach generalizes well to another task or not. 
In other words, it is not known which SLI model can be used for which contexts, tasks, and datasets.

To fill this gap, we conduct an extensive empirical evaluation of representative SLI models. 
To this end, we have for the first time defined and implemented the features of existing SLI models in a common framework available on GitHub.
Based on the discussion of related works, we chose six SLI models with unique feature combinations, which were developed for different tasks, to understand and compare their behavior for the data search task. 
We empirically investigate the behavior of each selected SLI model in three variants, where we index sub-graphs of $0$, $1$, and $2$ hop lengths.

The empirical evaluation consists of two sets of experiments.
In a first set of experiments, we analyze the relative size of the computed SLI compared to the original dataset (compression ratio) and the number of schema elements in the SLI compared to the number of data instances in the dataset (summarization ratio).
The second set of experiments quantifies the quality of a stream-based computation of the SLI for large datasets obtained from the Web of Data.
The stream-based approach is designed to scale to graphs of arbitrary sizes by observing the graph over a stream of edges with fixed window size.
Inherently, this approach introduces inaccuracies in the SLI computation by potentially extracting incomplete schema structures due to limited window size~\cite{SchemEX2012}.
Our experiments show huge variations in compression ratio, summarization ratio, and approximation quality for the SLI models.
However, we also observe strong positive and negative correlations between the three metrics. 
These insights shed light on the behaviors of SLI models for different datasets and queries that help to determine the right SLI model for a given task and dataset.

The remainder of this paper is structured as follows.
Subsequently, we discuss the features of the schema-level index (SLI) models reported in the literature. 
We formalize how data instances are summarized by SLIs in \cref{sec:schema-level-index}.
In \cref{sec:eval}, we introduce the experimental apparatus, \ie our framework and the datasets.
The results of our experiments are presented in \cref{sec:empirical-eval-1,sec:empirical-eval-2}.
Finally, we discuss our main findings, before we conclude.

\section{Discussion of Schema-level Index Models and Features}
\label{sec:relatedWork}
Our discussion of related works focuses on different models for schema-level indices (SLI).
As introduced above, SLIs support the execution of structural queries over the Web of Data.
Structural queries such as the example in \cref{lst:example-query} consist of a combination of types and properties.

Various SLI models were defined, which capture different schema structures and are defined using different theoretical approaches~\cite{DBLP:journals/vldb/CebiricGKKMTZ19}.
\cref{tab:index-model-features} is an overview of SLI models reported in the literature and the specific features they support.
%
%As shown in the table, we have identified ten features that are used (or explicitly not used) to compute the schema in existing index models.
These features are the use of property sets~\cite{CharacteristicSets:2011,SemSets:2012,SchemEX2012,DBLP:conf/esws/SpahiuPPRM16a,DBLP:conf/webi/BenedettiBP15,DBLP:conf/semweb/Mihindukulasooriya15,DBLP:conf/esws/SchaibleGS16,DBLP:conf/edbt/GoasdoueGM19}, use of type sets~\cite{SchemEX2012,DBLP:conf/esws/SpahiuPPRM16a,DBLP:conf/webi/BenedettiBP15,DBLP:conf/semweb/Mihindukulasooriya15,DBLP:conf/esws/SchaibleGS16,DBLP:conf/edbt/GoasdoueGM19}, use of neighbor information ~\cite{SemSets:2012,SchemEX2012,DBLP:conf/esws/SpahiuPPRM16a,DBLP:conf/webi/BenedettiBP15,DBLP:conf/semweb/Mihindukulasooriya15,DBLP:conf/esws/SchaibleGS16,DBLP:conf/edbt/GoasdoueGM19}, 
use of path information~\cite{CharacteristicSets:2011,SemSets:2012,SchemEX2012,DBLP:conf/esws/SpahiuPPRM16a,DBLP:conf/webi/BenedettiBP15,DBLP:conf/semweb/Mihindukulasooriya15,DBLP:conf/edbt/GoasdoueGM19}, 
use of $k$-bisimulation~\cite{DBLP:conf/edbt/GoasdoueGM19},
use of incoming property sets~\cite{CharacteristicSets:2011,DBLP:conf/edbt/GoasdoueGM19},
use of OR combination of feature~\cite{DBLP:conf/edbt/GoasdoueGM19},
use of transitively co-occurring (related) properties~\cite{DBLP:conf/edbt/GoasdoueGM19},
and use of inferred information from RDF Schema properties~\cite{DBLP:conf/esws/SpahiuPPRM16a,DBLP:conf/edbt/GoasdoueGM19}.
In addition, we propose inferencing over \texttt{owl:sameAs} as a new feature. 
\begin{table*}[!t]
\caption{\label{tab:index-model-features}Nine index models (left column) and their features (top row). Features marked with X are fully supported, (X) are partially supported, and - are not supported.}
\setlength{\tabcolsep}{4pt}
{\footnotesize
\begin{tabularx}{\linewidth}{p{4.5cm} *{10}{p{0.45cm}}}

\toprule
\backslashbox{\textbf{Index Model}}{\textbf{Feature}} & 
\rotateHeading{Property sets} & \rotateHeading{Type sets} & \rotateHeading{Neighbor information} & \rotateHeading{Path information} & \rotateHeading{k-bisimulation} & \rotateHeading{Incoming property set} &
\rotateHeading{OR combination} & \rotateHeading{Related properties} &  \rotateHeading{RDF Schema} & \rotateHeading{SameAs}  \\

\arrayrulecolor{black!100}\midrule
Characteristic Sets~\cite{CharacteristicSets:2011} & X & - & - & X & - & X & - & - & - & -\\
SemSets~\cite{SemSets:2012} & X & \hl{-} & \hl{X} & X & - & - & - & - & - & - \\
Weak Property Clique~\cite{DBLP:conf/edbt/GoasdoueGM19} &X& - & - & X & X & X & X & X & X & -\\
ABSTAT~\cite{DBLP:conf/esws/SpahiuPPRM16a} & X & X & X & X & - & - & - & - & (X) & - \\
LODex~\cite{DBLP:conf/webi/BenedettiBP15} & X & X & X & X & - & - & - & - & - & - \\
Loupe~\cite{DBLP:conf/semweb/Mihindukulasooriya15} & X & X & X & X & - & - & - & - & - & -  \\
SchemEX~\cite{SchemEX2012} & X & X & X & X & - & - & - & - & - & -  \\
TermPicker~\cite{DBLP:conf/esws/SchaibleGS16} & X & X & X & \hl{-} & - & - & - & - & - & - \\
\schemnew{}&X  & X & X & X & - & - & - & - & X & X \\

\arrayrulecolor{black!100}\bottomrule

\end{tabularx}
}
\end{table*}
Below, we discuss the SLI models presented in \cref{tab:index-model-features} from top to bottom.
We present their task, schema structure, and feature combination.

\begin{figure}[t!]
\centering
     \subfloat[\label{fig:charsets-example}Characteristic Sets~\cite{CharacteristicSets:2011} summarize data instances based on a common set of incoming and outgoing properties.]{%
       \includegraphics[trim={-2.0cm 0cm -2.0cm 0cm},clip,width=0.48\linewidth]{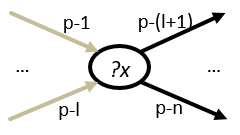}
     }
     \hfill
     \subfloat[\label{fig:semsets-example}SemSets~\cite{SemSets:2012} summarize data instances based on a common set of properties linked to the same resources.]{%
       \includegraphics[trim={-2.5cm 0cm -2.5cm 0cm},clip,width=0.48\linewidth]{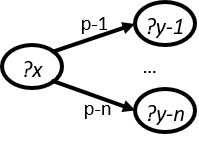}
     }
     \hfill
     \subfloat[\label{fig:weak-equivalence-example}Weak Property Cliques~\cite{DBLP:conf/edbt/GoasdoueGM19} summarize data instances based on at least one related incoming property and/or at least one related outgoing property.]{%
       \includegraphics[trim={-2.15cm 0cm -2.15cm 0cm},clip,width=0.48\linewidth]{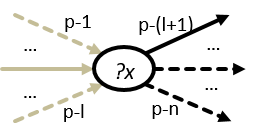}
     }
     \hfill
     \subfloat[\label{fig:schemex-example}SchemEX~\cite{SchemEX2012}, ABSTAT~\cite{DBLP:conf/esws/SpahiuPPRM16a}, LODeX~\cite{DBLP:conf/webi/BenedettiBP15}, and Loupe~\cite{DBLP:conf/semweb/Mihindukulasooriya15} summarize data instances based on a common type set and a common set of properties linked to resources sharing the same type set.]{%
       \includegraphics[trim={0cm 0cm 0cm 0cm},clip,width=0.48\linewidth]{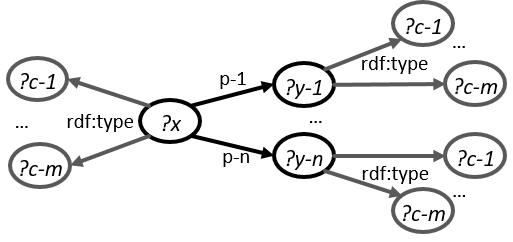}
     }
      \hfill
      \subfloat[\label{fig:TermPicker-example}TermPicker~\cite{DBLP:conf/esws/SchaibleGS16} summarizes data instances based on a common type set, a common property set, and a common type set linked over all properties.]{%
       \includegraphics[trim={-4cm 0cm -4cm 0cm},clip,width=1\linewidth]{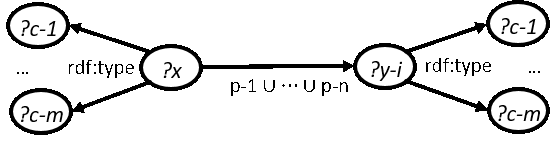}
     }
\caption{\label{fig:indices-examples}Schema structures captured by different index models from the related work.}
\end{figure}

Characteristic Sets~\cite{CharacteristicSets:2011}, illustrated in \cref{fig:charsets-example}, was developed to optimize cardinality estimations for join-queries in RDF databases.
Formally, the SLI model is defined as sets of data instances using a first-order-logic expression over triples.
Characteristic Sets summarize data instances along common incoming properties (the \textbf{incoming property set}) and outgoing properties (the \textbf{property set}).
Related to Characteristic Sets is the SLI model SemSets~\cite{SemSets:2012}, illustrated in \cref{fig:semsets-example}, which was developed to discover semantically similar sets of data instances in knowledge graphs to improve keyword-based ad-hoc retrieval.
SemSets is defined using set operators.
The SemSets model summarizes data instances that share the same outgoing properties, which are linked to a common resource. 
Thus, the model uses property sets and include \textbf{neighbor information}.

Goasdou{\'{e}} \etal~\cite{DBLP:conf/edbt/GoasdoueGM19} introduce compact graph summaries for RDF graphs.
They propose so-called Property Cliques formalized as equivalence relations, which can capture different schema structures.
Property Cliques summarize data instances based on the co-occurrence of \textbf{related properties}. 
In their work, two properties are related, if they co-occur in a data instance.
Thus, also pointed out by the authors, the relationship of being related is transitive~\cite{DBLP:conf/edbt/GoasdoueGM19}. 
Furthermore, Goasdou{\'{e}} \etal use property sets and incoming property sets. 
They propose so-called \enquote{strong equivalence} relations, where the related incoming \textit{and} outgoing properties are the same, and so-called \enquote{weak equivalence} relations, where the incoming \textit{or} the outgoing properties are the same (the \textbf{OR combination}).
They also support \textbf{RDF Schema} inferencing to compute their summaries.
The weak equivalence version of the Property Cliques is the only index model that uses transitively co-occurring (related) properties and summarizes instances based on the same property sets, or the same incoming property sets, or both (OR combination). 
In the following, we call this variant of the compact graph summaries \enquote{Weak Property Clique} (illustrated in \cref{fig:weak-equivalence-example}).

ABSTAT~\cite{DBLP:conf/esws/SpahiuPPRM16a}, LODeX~\cite{DBLP:conf/webi/BenedettiBP15}, Loupe~\cite{DBLP:conf/semweb/Mihindukulasooriya15}, and SchemEX~\cite{SchemEX2012} summarize data instances based on a common set of RDF types (the \textbf{type sets}) and properties linking to resources with the same type set (illustrated in \cref{fig:schemex-example}).
Thus, they use neighbor information in combination with type sets.
ABSTAT, LODeX, and Loupe were developed to explore datasets.
ABSTAT's schema structure has no formal definition and is only informally defined in a textual description.
In terms of the SLI model features, ABSTAT~\cite{DBLP:conf/esws/SpahiuPPRM16a} can use inferencing over the RDFS type hierarchies to compute so-called minimal patterns.
A minimal pattern is the set of most specific RDF types associated with the instances.
All super-types of these most specific types are removed.
This feature is also supported by Goasdou{\'{e}} \etal~\cite{DBLP:conf/edbt/GoasdoueGM19}, who additionally take sub-properties as well as RDFS domain and range into account.
Similarly, the SLI model Loupe is also only informally specified, which---for some aspects---leaves room for interpretations.
The Loupe model supports three so-called Class, Property, and Triple Inspector that allow different types of queries.
The Class Inspector considers only type sets, the Property Inspector only property sets, and the Triple Inspector a combination of type sets and property sets.
The Triple Inspector extracts triples of the form $<subjectType, predicate, objectType>$ from the input graph. 
Thus, the schema structure captured by Triple Inspector is equivalent to the SLI models ABSTAT and LODeX.
In contrast to ABSTAT and Loupe, the SLI model of LODeX computes clusters over the RDF types and selects a representative RDF type per cluster.
SchemEX was evaluated for the data search task on snapshots of the Web of Data using a stream-based schema extraction approach.
SchemEX was defined as stratified $1-$bisimulation~\cite{Bisimulation:2009}.

TermPicker~\cite{DBLP:conf/esws/SchaibleGS16}, illustrated in \cref{fig:TermPicker-example}, 
was developed to make data-driven recommendations of vocabulary terms.
TermPicker summarizes data instances based on a common type set, a common property set, and a common type set of all property-linked resources.
Thus, in contrast to, \eg SchemEX, TermPicker does not take \textbf{path information} into account, \ie it does not matter via which properties these data instances are connected.
Therefore, in order to model TermPicker one needs to be able to aggregate the information of neighboring data instances.

Furthermore, we propose with the index model \schemnew{} (short for: \schemnewLong{}) also a new combination of features. 
\schemnew{} extents the common schema structure of ABSTAT, LODeX, Loupe, and SchemEX, by providing support for inferencing over \textbf{\texttt{owl:sameAs}} as well as using all RDFS information, \ie types, properties, domain, and range. 

\textbf{Bisimulation} is an approach to determine whether two graphs are equivalent by considering the traversal over a graph as operating on state transition systems and defines an equivalence relation over states~\cite{Bisimulation:2009}. 
Two states are equivalent (or bisimilar) if they change into equivalent states with the same type of transition.
A stratified $k$-bisimulation reduces the path length to $k$ hops as it is defined with $k=1$ in SchemEX~\cite{SchemEX2012}.
Tran \etal~\cite{Tran2013} model $k$-bisimulation as height parameterization that can be applied on SLI.
Thus, we decided to use $k$-bisimulation as a feature rather than a separate index model.
Tran \etal determine that the parameter is typically $k \in {0, 1, 2}$.

In summary, from the discussion of the related works, we can state that there exists a variety of SLI models that capture different schema structures and are suitable for different tasks. 
SLI models are designed, implemented, and evaluated independently for their specific tasks.
There has never been a systematic comparison of SLI models nor has it been investigated how different SLI models, queries, and datasets influence the results of querying on the Web of Data.

\section{Summarizing Data Instances with Schema-level Indices}
\label{sec:schema-level-index}
We define the foundational concepts of data graph, data instances, and summarization of instances using schema-level indices (SLI). 
A \textbf{RDF data graph} $G$ is defined as $G \subseteq V_{U} \cup V_{B} \times P \times (V_{U} \cup V_{B} \cup L)$, where $V_{U}$ denotes the set of URIs, $V_{U}$ the set of blank nodes, $P \subseteq V_{U}$ the set of properties, and $L$ the set of literals.
Furthermore, there exists a subset $V_C \subseteq V_{U} \cup V_{B}$ that contains all RDF classes. 
A triple is a statement about a resource $s \in V_{U} \cup V_{B}$ in the form of a subject-predicate-object expression $(s,p,o) \in G$.

\begin{figure*}[!tb]
\centering
\includegraphics[trim={0cm 0cm 0cm 0cm},width=0.9\linewidth]{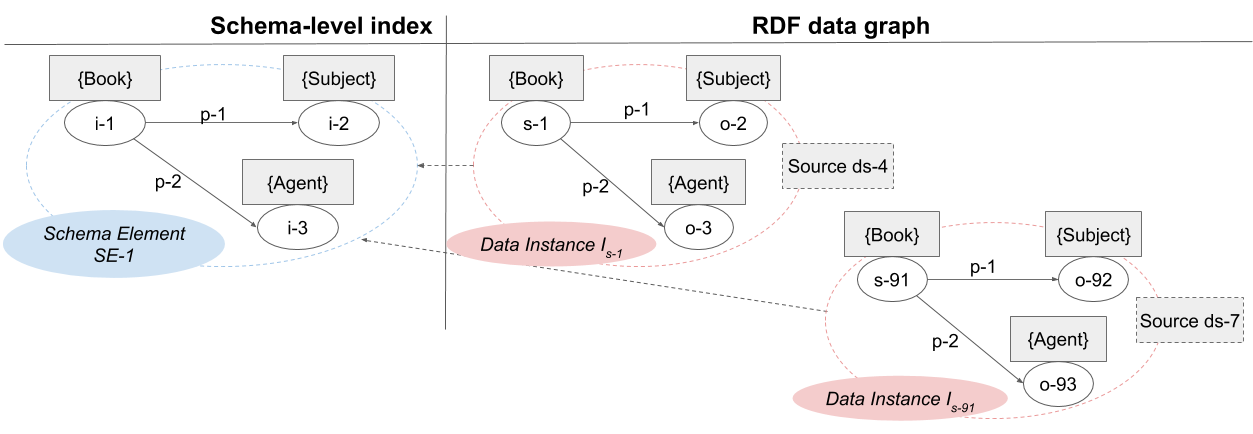}
\caption{\label{fig:SLI-example}Computing the SchemEX index for a decentralized RDF data graph. Both data instances $I_{s\mhyphen1}$ and $I_{s\mhyphen91}$ are summarized by the same schema element $SE\mhyphen1$.}
\end{figure*}

We define a \textbf{data instance} $I_s \subseteq G$ to be a set of triples, where each triple shares a common subject URI $s$.
We call the set of all $c \in V_C$ with $(s, \textrm{\textit{rdf:type}}, c) \in I_s$ the type set of $I_s$. 
Furthermore, we call the set of all $p \in P$ with $(s,p,o) \in I_s$ the property set of $I_s$. 
In our example in \cref{fig:SLI-example}, we visualize the type sets of the instances as squares above the corresponding URI. 
The remaining triples are visualized as directed edge, \eg $(s\mhyphen1,p\mhyphen1,o\mhyphen1)$.

SLI summarize data instances $I_s$ based on their common schema.
The specific features used to compute the schema of a data instance are defined in the SLI model.
For example, Characteristic Sets uses property sets and incoming property sets of each instance only, while most other index models also support to take RDF types into account.
For a detailed discussion of the features in SLI models, please refer to \cref{sec:relatedWork}.
Data instances with the same schema can be uniquely identified by their common schema.
This unique schema identifier, which summarizes the data instances, is called the \textbf{schema element}.
Schema elements are stored as keys in the SLI and can be used, \eg to retrieve all summarized data instances.
In \cref{fig:SLI-example}, the schema-level index contains one schema element (SE-1).

For a task such as searching in the Web of Data, one is interested in the data sources of the summarized instances (compare the motivating illustration in \cref{fig:SLI-setting}).
To model where data instances have been found on the Web of Data, we define the data source $D_s$ of an instance $I_s$ as the set of URIs $ds$ that identify documents on the Web that contain triple $(s,p,o)$ about instance $I_s$.
Common RDF Crawlers provide this information using the notion of quads $(s,p,o,ds)$~\cite{DBLP:conf/semweb/IseleUBH10}.
For data search, this means that the SLI does not need to store URIs of the summarized data instances directly, but merely needs to store the information about the URIs to the data sources where these instances have been observed.
For example, the data instances in \cref{fig:SLI-example} can be found in $ds\mhyphen4$ and $ds\mhyphen7$.
Thus, intuitively, a SLI for data search is a lookup table that contains information about schema structures of data instances on the Web and where they are located.
For another task such as cardinality estimation one would lookup the number of summarized data instances instead of their location.

In the Web of Data exists the idea of inferencing implicit information, \eg using RDFS or \texttt{owl:sameAs}. 
For example, the semantics of a triple\break $(p\mhyphen1, \textrm{\textit{rdfs:subPropertyOf}}, p\mhyphen2) \in G$ states that for any instance $I_s$ with $(s,p\mhyphen1,o)$, we can infer the additional triple $(s,p\mhyphen2,o)$. 
In the following, we assume that inferencing over RDF or \texttt{owl:sameAs} adds the respective types and properties to the corresponding type sets and property sets of each instance.

\section{Experimental Study}
\label{sec:eval}
We implement all features discussed in \cref{sec:relatedWork} into a single framework. 
Our framework allows flexibly combining these features to define SLI models and is available on GitHub along with descriptive dataset statistics, queries, and raw results (\url{https://github.com/t-blume/fluid-framework}).

\paragraph{\textbf{Choice of SLI Models.}} 
We select six representative index models based on our analysis in \cref{sec:relatedWork}.
These SLI models are Characteristic Sets, Weak Property Clique, SemSets, TermPicker, SchemEX, and \schemnew{}.
We select Characteristic Sets, Weak Property Clique, SemSets, TermPicker since they provide unique feature combinations.
Furthermore, we select SchemEX since the SLI model shares the same base schema structure with LODex, Loupe, and ABSTAT (see \cref{tab:index-model-features}).
Finally, we select \schemnew{} since it uses \texttt{owl:sameAs} and full RDFS reasoning.
This covers also the RDFS type hierarchy inferencing of ABSTAT and the RDFS reasoning provided by the compact graph summaries.
For each of the six selected SLI models, we apply the $k$-height parameterization feature proposed by Tran \etal~\cite{Tran2013}.
Reasonable values for the height parameterization $k$ are $0$, $1$, and $2$~\cite{Tran2013}.
Thus, in total, we compare 18 unique index models.

\paragraph{\textbf{Datasets.}} 
We use two datasets crawled from the Web of Data.
% 
% -- TimBL11M
The first dataset is called \emph{TimBL-11M}.
It contains about $11$ million triples crawls in a breadth-first search starting from a single seed URI, the FOAF profile of Tim Berners-Lee~\cite{SchemEX2012}. 
The TimBL-11M dataset is a graph with $673k$ data instances distributed over $18k$ data sources.
Each instance has on average $13$ outgoing properties (standard deviation (SD): $275$), $3$ incoming properties (SD: $848$), $3$ types (SD: $40$), and is defined in $3$ data sources (SD: $40$).
The latter means that, on overage, triples with a common subject URI can be found in $3$ distinct data sources, which makes the challenge for data search specifically hard.
Overall, $3,919$ unique properties and $2,738$ unique RDF types appear in the TimBL-11M dataset.

The second dataset \emph{DyLDO-127M} contains $127M$ triples~\cite{Kaefer2013}.
The Dynamic Linked Data Observatory (DyLDO) provides regular snapshots from the Web of Data.
In contrast to TimBL-11M, which uses only a single seed URI, the DyLDO dataset is populated by crawling in breadth-first search about $95k$ representative seed URIs of the Web of Data~\cite{Kaefer2013}.
The crawling of these seed URIs is restricted to a crawling depth of two hops.
We use the first snapshot of the DyLDO dataset, as it is with $127M$ triples the larged one.
The DyLDO-127M dataset contains $7M$ data instances distributed over $154k$ data sources.
Each instance has on average $17$ outgoing properties (SD: $6503$), $7$ incoming properties (SD: $635$), a single type (SD: $17$), and is defined in $2$ data sources (SD: $17$). 
Overall, $15k$ unique properties and $31k$ unique types appear in the DyLDO-127M dataset.

% Final comment
Both datasets are reasonably large with $11M$ triples and $127M$ triples, while still allowing to compute a gold standard for the stream-based index computation (\cref{sec:empirical-eval-2}).
A gold standard is created by loading the entire dataset into the main memory and computing the indices with no window size limit.
No pre-processing was conducted on the datasets, except removing triples that do not follow the W3C standards.

\begin{table*}[!t]
\caption{Results from the analysis of the \textbf{compression ratio} and \textbf{summarization ratio} of the six selected SLI models (with height parameter $k \in \lbrace 0,1,2\rbrace$).
$\#t$ is the number of triples in millions (M) in the SLI and in brackets below the ratio compared to the number of triples in the dataset (\textbf{compression ratio}).  
$\#e$ is the number of schema elements in thousands (T) in the SLI and in brackets below the ratio compared to the number of instances in the dataset (\textbf{summarization ratio}).
As datasets, we use the TimBL-11M (top) and DyLDO-127M datasets (bottom).}
 
{\renewcommand\arraystretch{1.1}
\setlength{\tabcolsep}{10pt}
%\small
\scriptsize
%\rotatebox[origin=b]{90}{%
\begin{tabularx}{1\textwidth}{p{0.05cm} p{2cm} *{3}{@{\hskip0.35in}c@{\hskip 0.06in}c}}
\arrayrulecolor{black!100}\toprule

& & \multicolumn{2}{@{}c@{\hskip0.35in}}{\textbf{$\bm{k} = 0$}}
& \multicolumn{2}{@{}c@{\hskip0.35in}}{\textbf{$\bm{k} = 1$}} 
& \multicolumn{2}{@{}c@{\hskip0.35in}}{\textbf{$\bm{k} = 2$}}\\

& \textbf{Index Model}  & \textbf{$\bm{\#t}$} & \textbf{$\bm{\#e}$} & \textbf{$\bm{\#t}$} & \textbf{$\bm{\#e}$} & \textbf{$\bm{\#t}$} & \textbf{$\bm{\#e}$} \\

\arrayrulecolor{black!100}\midrule
\parbox[t]{4mm}{\multirow{12}{*}{\rotatebox[origin=c]{90}{\textbf{\shortstack[c]{TimBL-11M}}}}}

& \multirow{2}{*}{\textbf{\shortstack{Character\\-istic Sets}}} & \multirow{2}{*}{\shortstack{$na$\\ $(na)$}} & \multirow{2}{*}{\shortstack{$na$\\ $(na)$}} & \multirow{2}{*}{\shortstack{$0.7M$\\ $(6.5\%)$}} & \multirow{2}{*}{\shortstack{$9.6T$\\ $(1.4\%)$}} & \multirow{2}{*}{\shortstack{$1.6M$\\ $(14.6\%)$}} & \multirow{2}{*}{\shortstack{$37.2T$\\ $(5.5\%)$}}\\

\arrayrulecolor{black!30}\cmidrule(r{1pt}){2-8}

& \multirow{2}{*}{\textbf{\shortstack{Weak Prop\\-erty Clique}}} & \multirow{2}{*}{\shortstack{$na$\\ $(na)$}} & \multirow{2}{*}{\shortstack{$na$\\ $(na)$}} & \multirow{2}{*}{\shortstack{$1.9M$\\ $(17.9\%)$}} & \multirow{2}{*}{\shortstack{$74$\\ $(< 0.1\%)$}} & \multirow{2}{*}{\shortstack{$1.1M$\\ $(9.9\%)$}} & \multirow{2}{*}{\shortstack{$50$\\ $(< 0.1\%)$}}\\
\\
\arrayrulecolor{black!30}\cmidrule(r{1pt}){2-8}

& \textbf{SemSets} & \multirow{2}{*}{\shortstack{$0.3M$\\$(2.9\%)$}} & \multirow{2}{*}{\shortstack{$2.8T$\\ $(0.4\%)$}} & \multirow{2}{*}{\shortstack{$7.6M$\\ $(69.2\%)$}} & \multirow{2}{*}{\shortstack{$139.0T$\\ $(20.6\%)$}} & \multirow{2}{*}{\shortstack{$7.6M$\\ $(69.2\%)$}} & \multirow{2}{*}{\shortstack{$139.0T$\\ $(69.2\%)$}}\\
\\
\arrayrulecolor{black!30}\cmidrule(r{1pt}){2-8}

& \textbf{SchemEX} & \multirow{2}{*}{\shortstack{$0.3M$\\$(2.9\%)$}} & \multirow{2}{*}{\shortstack{$2.8T$\\ $(0.4\%)$}} & \multirow{2}{*}{\shortstack{$0.8M$\\ $(6.9\%)$}} & \multirow{2}{*}{\shortstack{$12.0T$\\ $(1.8\%)$}} & \multirow{2}{*}{\shortstack{$1.4M$\\ $(12.5\%)$}} & \multirow{2}{*}{\shortstack{$27.7T$\\ $(4.1\%)$}}\\
\\
\arrayrulecolor{black!30}\cmidrule(r{1pt}){2-8}

& \textbf{TermPicker} & \multirow{2}{*}{\shortstack{$0.3M$\\$(2.9\%)$}} & \multirow{2}{*}{\shortstack{$2.8T$\\ $(0.4\%)$}} & \multirow{2}{*}{\shortstack{$0.7M$\\ $(6.5\%)$}} & \multirow{2}{*}{\shortstack{$10.8T$\\ $(1.6\%)$}} & \multirow{2}{*}{\shortstack{$1.8M$\\ $(16.0\%)$}} & \multirow{2}{*}{\shortstack{$37.3T$\\ $(5.5\%)$}}\\
\\
\arrayrulecolor{black!30}\cmidrule(r{1pt}){2-8}
& \multirow{2}{*}{\textbf{\shortstack{SchemEX\\+U+I}}} & \multirow{2}{*}{\shortstack{$0.4M$\\$(3.8\%)$}} & \multirow{2}{*}{\shortstack{$3.1T$\\ $(0.5\%)$}} & \multirow{2}{*}{\shortstack{$0.8M$\\ $(7.1\%)$}} & \multirow{2}{*}{\shortstack{$11.3T$\\ $(1.7\%)$}} & \multirow{2}{*}{\shortstack{$1.8M$\\ $(15.9\%)$}} & \multirow{2}{*}{\shortstack{$31.0T$\\ $(4.6\%)$}}\\
\\
\arrayrulecolor{black!100}\midrule

\parbox[t]{4mm}{\multirow{14}{*}{\rotatebox[origin=c]{90}{\textbf{\shortstack[c]{DyLDO-127M}}}}}

& \multirow{2}{*}{\textbf{\shortstack{Character\\-istic Sets}}} & \multirow{2}{*}{\shortstack{$na$\\ $(na)$}} & \multirow{2}{*}{\shortstack{$na$\\ $(na)$}} & \multirow{2}{*}{\shortstack{$0.6M$\\ $(0.5\%)$}} & \multirow{2}{*}{\shortstack{$23.0T$\\ $(0.3\%)$}} & \multirow{2}{*}{\shortstack{$2.1M$\\ $(1.7\%)$}} & \multirow{2}{*}{\shortstack{$112.8T$\\ $(1.6\%)$}}\\

\arrayrulecolor{black!30}\cmidrule(r{1pt}){2-8}

& \multirow{2}{*}{\textbf{\shortstack{Weak Prop\\-erty Clique}}} & \multirow{2}{*}{\shortstack{$na$\\ $(na)$}} & \multirow{2}{*}{\shortstack{$na$\\ $(na)$}} & \multirow{2}{*}{\shortstack{$14.8M$\\ $(9.0\%)$}} & \multirow{2}{*}{\shortstack{$394$\\ $(< 0.1\%)$}} & \multirow{2}{*}{\shortstack{$25.1M$\\ $(19.7\%)$}} & \multirow{2}{*}{\shortstack{$102$\\ $(< 0.1\%)$}}\\
\\
\arrayrulecolor{black!30}\cmidrule(r{1pt}){2-8}

& \textbf{SemSets} & \multirow{2}{*}{\shortstack{$4.1M$\\$(3.2\%)$}} & \multirow{2}{*}{\shortstack{$46.6T$\\ $(0.7\%)$}} & \multirow{2}{*}{\shortstack{$45.3M$\\ $(35.6\%)$}} & \multirow{2}{*}{\shortstack{$1733.5T$\\ $(25.0\%)$}} & \multirow{2}{*}{\shortstack{$45.3M$\\ $(35.6\%)$}} & \multirow{2}{*}{\shortstack{$1733.5T$\\ $(25.0\%)$}}\\
\\
\arrayrulecolor{black!30}\cmidrule(r{1pt}){2-8}

& \textbf{SchemEX} & \multirow{2}{*}{\shortstack{$4.1M$\\$(3.2\%)$}} & \multirow{2}{*}{\shortstack{$46.6T$\\ $(0.7\%)$}} & \multirow{2}{*}{\shortstack{$15.7M$\\ $(12.3\%)$}} & \multirow{2}{*}{\shortstack{$254.5T$\\ $(3.6\%)$}} & \multirow{2}{*}{\shortstack{$19.8M$\\ $(15.6\%)$}} & \multirow{2}{*}{\shortstack{$431.1T$\\ $(6.1\%)$}}\\
\\
\arrayrulecolor{black!30}\cmidrule(r{1pt}){2-8}

& \textbf{TermPicker} & \multirow{2}{*}{\shortstack{$4.1M$\\$(3.2\%)$}} & \multirow{2}{*}{\shortstack{$46.6T$\\ $(0.7\%)$}} & \multirow{2}{*}{\shortstack{$11.1M$\\ $(8.7\%)$}} & \multirow{2}{*}{\shortstack{$238.4T$\\ $(3.4\%)$}} & \multirow{2}{*}{\shortstack{$25.4M$\\ $(19.9\%)$}} & \multirow{2}{*}{\shortstack{$559.1T$\\ $(7.9\%)$}}\\
\\
\arrayrulecolor{black!30}\cmidrule(r{1pt}){2-8}
& \multirow{2}{*}{\textbf{\shortstack{SchemEX\\+U+I}}} & \multirow{2}{*}{\shortstack{$8.5M$\\$(6.7\%)$}} & \multirow{2}{*}{\shortstack{$53.0T$\\ $(0.8\%)$}} & \multirow{2}{*}{\shortstack{$19.9M$\\ $(15.7\%)$}} & \multirow{2}{*}{\shortstack{$249.5T$\\ $(3.5\%)$}} & \multirow{2}{*}{\shortstack{$22.9M$\\ $(18.0\%)$}} & \multirow{2}{*}{\shortstack{$466.9T$\\ $(6.6\%)$}}\\
\\
\arrayrulecolor{black!100}\bottomrule
\end{tabularx}
%}
\label{tab:index-compression}\label{tab:index-summarization}
}
\end{table*}

\section{Experiment 1: Compression and Summarization Ratio}
\label{sec:empirical-eval-1}

\paragraph{\textbf{Metrics.}} 
We evaluate the index size for the selected indices over the two datasets mentioned above. 
The size of an index refers to the number of triples when stored as an RDF graph.
We compare the number of triples in the index to the number of triples in the dataset (\textbf{compression ratio}).
Furthermore, we compare the number of schema elements in the index to the number of data instances in the dataset (\textbf{summarization ratio}). 
This ratio gives an idea of how well the defined schema structure can summarize data instances on the Web of Data.
For the compression and summarization ratios, we use exact indices.
This means we loaded the complete data graph into the main memory before we started the index computation process.

\paragraph{\textbf{Results.}} 
The results of the experiments regarding the compression ratio and summarization ratio are documented in \cref{tab:index-summarization}. 
As one can see, there is a huge variety in terms of how well indices compress and summarize the data.
For the TimBL-11M dataset, SemSets' compression ratio (with $k = 1$) is about 10 times larger than all other indices except for Weak Property Cliques (only about 5 times larger).
For the DyLDO-127M dataset, SemSets' compression ratio (with $k = 1$) is up to 75 times larger.
Additionally, there is no increase in index size from $k = 1$ to $k = 2$, but a more than ten-times increase from $k = 0$ to $k = 1$.
A similar increase appears for the summarization ratio. 
SemSets is the only index that uses neighbor information but not neighbor type sets, \ie they compare the object URIs $o$ of each $(s,p,o)$ triple.
In contrast, the other indices either ignore objects or consider their type sets only.
SemSets has a summarization ratio of $20\% - 25\%$, \ie on average $4 - 5$ data instances share the same schema structure. 
The smallest index, Characteristic Sets, has a summarization ratio of $0.3\%$, \ie about $330$ data instances share the same schema structure.
A notable exception is the Weak Property Clique, which shows the most condensed summarization (summarization ratio of less than $0.1\%$).
However, the combination of either incoming or outgoing related properties in Weak Property Cliques leads to a considerably large compression ratio. 
Weak Property Clique indices are more than twice the size of Characteristic Sets indices.

When considering the semantics of RDFS and \textit{owl:sameAs} in SchemEX+U+I, the index size increases compared to SchemEX by about $3\%$ more triples.
Despite being a larger index in terms of the number of triples, for $k = 1$ fewer schema elements are computed when including the semantics of RDFS and \textit{owl:sameAs}.
For $k = 0$ and $k = 2$, SchemEX+U+I requires more schema elements than SchemEX to summarize the data instances.

In summary, including the semantics of \textit{owl:sameAs} and RDF Schema increases the size of the index.
However, it can reduce the number of schema elements.
Furthermore, using weak equivalences leads to a handful of schema elements with a considerably large size summarizing all data instances.

\section{Experiment 2: Stream-based Index Computation}
\label{sec:empirical-eval-2}
In this experiment, we are interested in how well queries of varying complexity can be supported by the indices if the SLI is computed over a stream of graph edges.
Motivated from stream-databases, the idea is to consider the triples in the datasets as a stream that is observed in windows of sizes $1k$, $100k$, and $200k$.
This allows us to scale the computation to in principle arbitrary sized input graphs~\cite{SchemEX2012}.
However, the approach produces approximation errors since only a fraction of the data graph is kept simultaneously in the main memory, while the remainder is not yet known or inaccessible.
Thus, we potentially extract incomplete schema structures.

Regarding the index model of \schemnew{}, we evaluate two variants in this experiment:
The RDFS inferencing requires an additional data structure during the computation process, the so-called schema graph~\cite{DBLP:journals/corr/abs-1908-01528}.
This schema graph is constructed from the triples using RDFS \texttt{range}, \texttt{domain}, \texttt{subClassOf}, or \texttt{subPropertyOf}.
With the domain, range, and hierarchical types/properties information, we infer additional types and properties for the remaining data instances.
In one version called \schemnewo{}, the RDFS information is extracted and inferred \emph{on-the-fly}.
Here, we construct the schema graph simultaneously to the index computation. 
The advantage is that only one pass over the dataset is needed.
However, since the schema graph is built while the index is computed information may be missing for the inferencing.
In the other version called \schemnewp{}, we first extract all RDFS information in a \emph{pre-processing} step to construct the schema graph.
The advantage is that the inferencing of triples is conducted on the complete schema graph only.
The drawback is that two passes over the dataset are needed.

\paragraph{\textbf{Queries.}} 
A central challenge for this experiment is the choice of queries to be executed over the indices.
Here, we follow the work by Konrath \etal~\cite{SchemEX2012} who conducted a data-driven query generation for the evaluation of approximate graph indices.
This means the queries are generated from the actual data instances in the datasets, \ie their combination of types and properties.
We distinguish two types of queries, simple queries (SQ) and complex queries (CQ).
Simple queries search for data instances that have a common type set (or in the case of SemSets a common set of objects).
Analyses of existing query logs show that most SPARQL queries in search systems are simple queries~\cite{DBLP:journals/corr/abs-1103-5043}.
In contrast, complex queries search for data instances that match the complete schema structure defined by the specific index model, \eg include property paths over $2$ hops for Characteristic Sets with $k=2$.

\paragraph{\textbf{Metric.}}
We execute the simple and complex queries on the SLI computed with fixed window size and on the gold standard SLI.
For our data search task, the results of the queries are the two sets $D_{gold}$ and $D_{window}$, which contain the corresponding data source URIs.
Following Konrath \etal~\cite{SchemEX2012}, the approximation quality is measured by comparing $D_{gold}$ and $D_{window}$ using the F1-score.

\begin{figure}[t!]
     \subfloat[F1-scores for TimBL-11M and $k=0$.\label{fig:timbl-k0}]{%
       \includegraphics[trim={0.25cm 0.25cm 0.25cm 1.75cm},clip,width=0.48\textwidth]{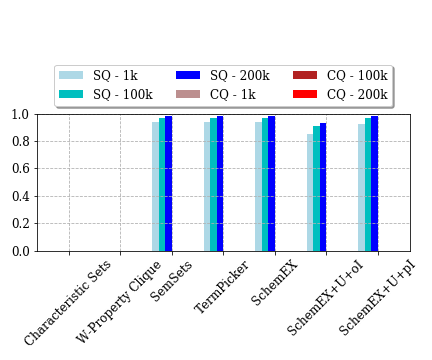}
     }
     \hfill
     \subfloat[F1-scores for DyLDO-127M and $k=0$.\label{fig:dyldo-k0}]{%
       \includegraphics[trim={0.25cm 0.25cm 0.25cm 1.75cm},clip,width=0.48\textwidth]{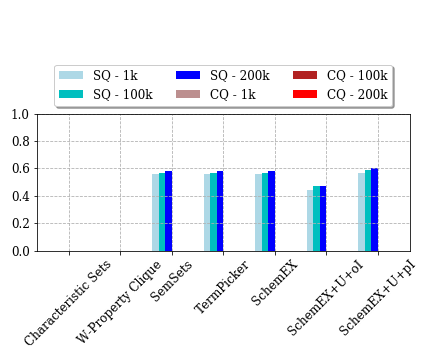}
     }
     \\
     \subfloat[F1-scores for TimBL-11M and $k=1$.\label{fig:timbl-k1}]{%
       \includegraphics[trim={0.25cm 0.25cm 0.25cm 3.86cm},clip,width=0.48\textwidth]{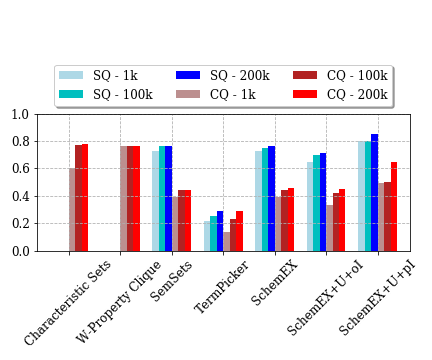}
     }
     \hfill
     \subfloat[F1-scores for DyLDO-127M and $k=1$.\label{fig:dyldo-k1}]{%
       \includegraphics[trim={0.25cm 0.25cm 0.25cm 3.86cm},clip,width=0.48\textwidth]{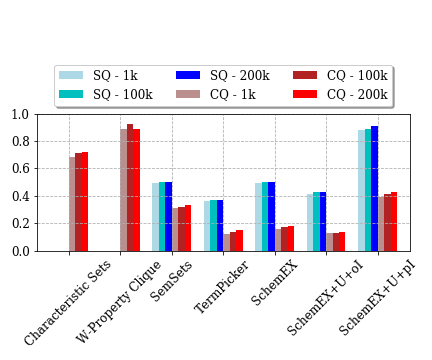}
     }
     \\
	 \subfloat[F1-scores for TimBL-11M and $k=2$.\label{fig:timbl-k2}]{%
       \includegraphics[trim={0.25cm 0.25cm 0.25cm 3.85101cm},clip,width=0.48\textwidth]{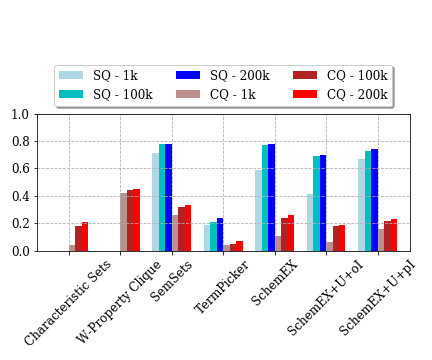}
     }
     \hfill
     \subfloat[F1-scores for DyLDO-127M and $k=2$.\label{fig:dyldo-k2}]{%
       \includegraphics[trim={0.25cm 0.25cm 0.25cm 3.85101cm},clip,width=0.48\textwidth]{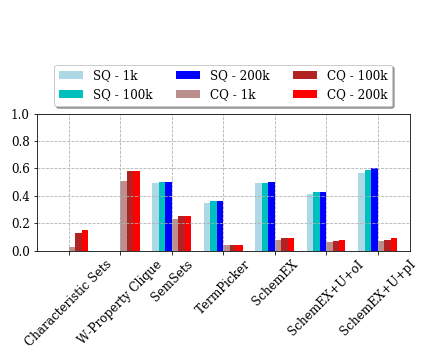}
     }
     \caption{F1-score for simple queries (SQ) and complex queries (CQ) and for window sizes (1k, 100k, 200k). The left column shows the values for the TimBL-11M dataset and the right column the DyLDO-127M dataset, respectively. The influence of the height parameter $k \in \lbrace 0,1,2\rbrace$ can be seen in the rows from top to bottom.}
     \label{fig:approximation-results}
\end{figure}

\paragraph{\textbf{Results.}}
\cref{fig:approximation-results} shows the approximation quality in terms of F1-score for the selected index models.
\cref{tab:results}, in the appendix, shows the same results but as a table.
For indices with a height parameter $k = 0$, the simple queries and the complex queries are alike.
Moreover, Characteristic Sets and Weak Property Cliques do not use type information (or object information).
Thus, simple queries are not available for these index models.

From the results of our experiment, we can state that simple queries consistently show higher F1-scores than complex queries. 
TermPicker and Weak Property Cliques are the only indices that have a higher F1-score on the DyLDO-127M dataset than on the TimBl-11M dataset.
As described in \cref{sec:relatedWork}, TermPicker is the only index not using the path information feature.
This restriction is the only difference in the schema structure compared to SchemEX.
Still, TermPicker has a $50\%$ lower F1-score than SchemEX on the TimBL-11M dataset.

Regarding the complex queries, the F1-scores of Weak Property Cliques are the highest in the experiments with $k = 1$ and $2$. 
For $k = 2$, Weak Property Cliques have between $.12$ and $.54$ higher F1-scores compared to the other indices.
SemSets only have a small drop in F1-score from $k = 1$ to $k = 2$.
\schemnewo{} consistently has a lower F1-score than SchemEX.
For\break \schemnewp{}, we extracted the RDF Schema information in a pre-processing step. 
Compared to \schemnewo{}, \schemnewp{} has consistently higher F1-scores.
Furthermore, for window sizes $100k$ and $200k$, \schemnewp{} has higher F1-scores than SchemEX (except for $k=2$ for TimBL-11M).

We also observe an influence of the characteristics of the crawled dataset on the approximation quality. 
All indices have on average a $.15$ lower F1-score on the DyLDO-127M dataset compared to the TimBL-11M dataset.
In particular, simple queries achieve much lower F1-scores.
On average, simple queries have $.25$ lower F1-scores and complex queries have $.04$ lower F1-scores on the DyLDO-127M dataset compared to the TimBL-11M dataset.
Furthermore, larger window sizes consistently improve F1-scores.
In contrast, on-the-fly inferencing lowered the F1-scores in our experiment compared to no inferencing.

\section{Discussion}
\label{sec:discussion}
Key insights from our experiments are:
(1) SLI models perform very differently in terms of compression ratio, summarization ratio, and approximation quality depending on the queries as well as the characteristic of the dataset.
(2) The approximation quality of an index computed in a stream-based approach depends on three factors:
First, we observe an influence of the characteristics of the crawled dataset. 
Second, simple queries consistently outperform complex queries. 
Third, a larger window size typically improves the quality only marginally. 

Regarding the first insight, we conducted a detailed analysis to understand the relationship between compression ratio and summarization ratio.
We computed the Pearson and Spearman correlation coefficient for the $n = 32$ SLI reported in \cref{tab:index-compression}. 
Results of the Pearson correlation indicated that there was a significant relationship between compression ratio and summarization ratio, 
$r(30)=.84, p < .0001$, and as well as for Spearman, $r_{s}(30)=.64, p < .0001$.
Furthermore, there is a significant negative correlation between summarization ratio and approximation quality of a stream-based computation approach.
We computed the Pearson and Spearman correlation coefficient for the three cache sizes $1k$, $100k$, and $200k$.
We compared the reported F1-scores for the complex queries (for $k = 0$, we used the simple queries) for each cache size (\cref{fig:approximation-results}) to the summarization ratio of the corresponding gold standard index (\cref{tab:index-compression}).
For all three cache sizes, we found a negative correlation coefficient with $p <.05$ (see \cref{tab:correlation}).
 
{\renewcommand\arraystretch{1.0}
\setlength{\tabcolsep}{6pt}
\begin{table}
%\scriptsize
\centering
\small
\caption{\label{tab:correlation}Results of correlation analysis between summarization ratio and approximation quality (F1-score). Pearson and Spearman coefficients and respected p-values for $n = 32$ SLI (\cref{tab:index-compression}) for three cache sizes with a degree of freedom $df = 30$.}
\begin{tabularx}{\columnwidth}{p{2cm} p{.75cm} p{1.5cm} p{.75cm} p{1.5cm} p{.75cm} p{2cm}}
\toprule
\textbf{Coefficient} &  \multicolumn{2}{c}{\textbf{1k}} & \multicolumn{2}{c}{\textbf{100k}} &\multicolumn{2}{c}{\textbf{200k}}	\\
\midrule

\textbf{Pearson} & $-.35$ & $p < .05$ & $-.38$ & $p < .04$ & $-.38$ & $p < .04$ \\

\textbf{Spearman}: & $-.74$ & $p < .0001$ & $-.74$ & $p < .0001$ & $-.75$ & $p < .0001$ \\

\bottomrule
\end{tabularx}
\end{table}
}

From the statistical analysis, we can see that a lower summarization ratio leads to a higher F1-score. 
This means index structures that summarize well, \ie summarize many data instances to the same schema element, can be computed with high accuracy in a stream-based approach.
When we compute correct schema elements in the stream-based approach, for index models with a low summarization ratio, we assign more data instances to the correct schema element than for index models with a high summarization ratio. 

Note that the extreme summarization ratio of Weak Property Cliques also produces the highest F1-scores.
This can point to a explanation for the observed correlation. 
With only a handful of schema elements in the index (see \cref{tab:index-compression}), it is more likely that a data instance is summarized by the correct schema element. 
This results in an overall higher F1-score.

We also observe an influence of the characteristics of how the data has been crawled.
First, all indices have on average a $.15$ lower F1-score on the DyLDO-127M dataset compared to the TimBL-11M dataset.
We explain this observation by the different crawling strategies.
The TimBL-11M dataset was crawled starting from a single seed URI.
In contrast, the DyLDO-127M dataset was crawled using more than $95,000$ seed URIs from $652$ unique pay-level domains~\cite{Kaefer2013}.
Furthermore, the crawling depth is limited to two hops.
Because of this difference in the crawling strategy, data instances bare different characteristics in both datasets.
First, the DyLDO-127M dataset contains nearly $4$-times more unique properties and about $11$-times more unique types as the TimBL-11M dataset (see \cref{sec:eval}).
Moreover, the TimBL-11M contains fewer data sources, and data instances are defined in fewer data sources than in the DyLDO-127M dataset.
This could be one possible explanation for the overall better performance on the TimBL-11M dataset.
The dataset characteristic also influences the size of the index. 
On average, the compression ratio of indices computed for the TimBL-11M dataset is $14.9\%$ and for the DyLDO-127M dataset, it is $10.5\%$.
Additionally, data instances in the DyLDO-127M dataset have more variety in the number of outgoing properties, but less variety in the number of types.
However, the indices using types (SchemEX, TermPicker, \schemnew{}) consistently achieve better compression and summarization ratios on the TimBL-11M dataset.
The evaluated indices not using types (Characteristic Sets, W-Property Cliques, SemSemts) achieve better compression and summarization ratios on the DyLDO-127M dataset.
Thus, the complexity of the combination of type sets and properties seems to be predominately impacted by the number of properties rather than the number of types.

Finally, we observe that inferencing RDF Schema information on-the-fly (\schemnewo{}) leads to lower F1-scores than inferencing in a pre-processing step (\schemnewp{}).
For \schemnewo{}, the schema graph information used for inferencing is incomplete until the last triple using an RDFS property is processed. 
Thus, for \schemnewo{} inferencing is another source for approximation errors.
However, while including the semantics of \texttt{owl:sameAs} and RDFS increases the size of the index, it reduced the number of schema elements in some experiments, \ie it achieves a better summarization ratio.

\section{Conclusion}
\label{sec:conclusion}
Our empirical evaluations reveal huge variations in compression ratio, summarization ratio, and approximation quality for different index models, queries, and datasets.
This confirms our hypothesis that there is no single schema-level index model that equally fits all tasks and that the performance of the SLI model depends on the specific types of queries and characteristics of the datasets.
However, we observed meaningful correlations in the results that help to determine the right index model for a given task, type of query, and dataset.
\emph{On GitHub, one can find our source code, the queries, and the detailed statistics about the TimBL-11M and DyLDO-127M datasets: \url{http://github.com/t-blume/fluid-framework}.}

\hfill \break
\noindent\textbf{Acknowledgment.}
This research was co-financed by the EU H2020 project MOVING (\url{http://www.moving-project.eu/}) under contract no 693092.

\bibliographystyle{splncs04}

\bibliography{bibliography}

\begin{thebibliography}{10}
\providecommand{\url}[1]{\texttt{#1}}
\providecommand{\urlprefix}{URL }
\providecommand{\doi}[1]{https://doi.org/#1}

\bibitem{DBLP:journals/corr/abs-1103-5043}
Arias, M., Fern{\'{a}}ndez, J.D., Mart{\'{\i}}nez{-}Prieto, M.A., de~la Fuente,
  P.: An empirical study of real-world {SPARQL} queries. CoRR
  \textbf{abs/1103.5043} (2011)

\bibitem{DBLP:conf/webi/BenedettiBP15}
Benedetti, F., Bergamaschi, S., Po, L.: Exposing the underlying schema of {LOD}
  sources. In: Joint {IEEE/WIC/ACM WI} and {IAT}. pp. 301--304. {IEEE} (2015)

\bibitem{DBLP:journals/corr/abs-1908-01528}
Blume, T., Scherp, A.: {FLuID}: {A} meta model to flexibly define schema-level
  indices for the web of data. CoRR  \textbf{abs/1908.01528} (2019)

\bibitem{DBLP:journals/vldb/CebiricGKKMTZ19}
{\v{C}}ebiric, {\v{S}}., Goasdou{\'{e}}, F., Kondylakis, H., Kotzinos, D.,
  Manolescu, I., Troullinou, G., Zneika, M.: Summarizing semantic graphs: a
  survey. {VLDB} J.  \textbf{28}(3),  295--327 (2019)

\bibitem{SemSets:2012}
Ciglan, M., N{\o}rv{\aa}g, K., Hluch{\'{y}}, L.: The {SemSets} model for ad-hoc
  semantic list search. In: WWW. pp. 131--140. {ACM} (2012)

\bibitem{DBLP:conf/edbt/GoasdoueGM19}
Goasdou{\'{e}}, F., Guzewicz, P., Manolescu, I.: Incremental structural
  summarization of {RDF} graphs. In: {EDBT}. pp. 566--569. OpenProceedings.org
  (2019)

\bibitem{Lodatio:2013}
Gottron, T., Scherp, A., Krayer, B., Peters, A.: {LODatio}: using a
  schema-level index to support users infinding relevant sources of {Linked
  Data}. In: {K-CAP}. pp. 105--108. {ACM} (2013)

\bibitem{DBLP:conf/rweb/HoseSTW11}
Hose, K., Schenkel, R., Theobald, M., Weikum, G.: Database foundations for
  scalable {RDF} processing. In: Reasoning Web. - Int. Summer School. LNCS,
  vol.~6848, pp. 202--249. Springer (2011)

\bibitem{DBLP:conf/semweb/IseleUBH10}
Isele, R., Umbrich, J., Bizer, C., Harth, A.: {LD}spider: An open-source
  crawling framework for the web of linked data. In: {ISWC} Posters {\&}
  Demonstrations. vol.~658. CEUR-WS.org (2010)

\bibitem{Kaefer2013}
K{\"{a}}fer, T., Abdelrahman, A., Umbrich, J., O'Byrne, P., Hogan, A.:
  Observing linked data dynamics. In: {ESWC}. vol.~7882, pp. 213--227. Springer
  (2013)

\bibitem{SchemEX2012}
Konrath, M., Gottron, T., Staab, S., Scherp, A.: {SchemEX} - efficient
  construction of a data catalogue by stream-based indexing of {Linked Data}.
  J. Web Sem.  \textbf{16},  52--58 (2012)

\bibitem{DBLP:conf/ekaw/LeiUM06}
Lei, Y., Uren, V.S., Motta, E.: {SemSearch}: {A} search engine for the semantic
  web. In: {EKAW}. vol.~4248, pp. 238--245. Springer (2006)

\bibitem{DBLP:conf/semweb/Mihindukulasooriya15}
Mihindukulasooriya, N., Poveda{-}Villal{\'{o}}n, M., Garc{\'{\i}}a{-}Castro,
  R., G{\'{o}}mez{-}P{\'{e}}rez, A.: Loupe - an online tool for inspecting
  datasets in the {Linked Data} cloud. In: ISWC Posters {\&} Demos. vol.~1486.
  CEUR-WS.org (2015)

\bibitem{CharacteristicSets:2011}
Neumann, T., Moerkotte, G.: Characteristic sets: Accurate cardinality
  estimation for {RDF} queries with multiple joins. In: {ICDE}. pp. 984--994.
  {IEEE} (2011)

\bibitem{pietriga2018browsing}
Pietriga, E., G{\"{o}}z{\"{u}}kan, H., Appert, C., Destandau, M., Cebiric, S.,
  Goasdou{\'{e}}, F., Manolescu, I.: Browsing linked data catalogs with
  {LODA}tlas. In: {ISWC}. vol. 11137, pp. 137--153. Springer (2018)

\bibitem{Bisimulation:2009}
Sangiorgi, D.: On the origins of bisimulation and coinduction. ACM Trans.
  Program. Lang. Syst.  \textbf{31}(4),  15:1--15:41 (2009)

\bibitem{DBLP:conf/esws/SchaibleGS16}
Schaible, J., Gottron, T., Scherp, A.: {TermPicker}: Enabling the reuse of
  vocabulary terms by exploiting data from the {Linked Open Data} cloud. In:
  {ESWC}. vol.~9678, pp. 101--117. Springer (2016)

\bibitem{DBLP:conf/esws/SpahiuPPRM16a}
Spahiu, B., Porrini, R., Palmonari, M., Rula, A., Maurino, A.: {ABSTAT:}
  ontology-driven linked data summaries with pattern minimalization. In: {ESWC}
  Satellite Events, Revised Selected Papers. vol.~9989, pp. 381--395 (2016)

\bibitem{Tran:2009:SSU:1615258.1615264}
Tran, T., Haase, P., Studer, R.: Semantic search - using graph-structured
  semantic models for supporting the search process. In: {ICCS}. pp. 48--65.
  Springer (2009)

\bibitem{Tran2013}
Tran, T., Ladwig, G., Rudolph, S.: Managing structured and semi-structured
  {RDF} data using structure indexes. {TKDE}  \textbf{25}(9),  2076--2089
  (2013)

\end{thebibliography}

\begin{landscape}
\appendix
\section*{Appendix}
\begin{table}[!th]
\caption{F1-score or not available (na) for simple queries (SQ) and complex queries (CQ), the height parameter\newline $k \in \lbrace 0,1,2\rbrace$, and for different window sizes (1k, 100k, 200k) on the two datasets TimBL-11M and DyLDO-127M.}
{\renewcommand\arraystretch{1.5}
\setlength{\tabcolsep}{4pt}
\centering

\scriptsize 
%\rotatebox[origin=b]{90}{%
\centering
\begin{tabularx}{1.35\textwidth}{lll *{7}{@{\hskip0.225in}c@{\hskip 0.06in}c@{\hskip 0.06in}c}}
    \toprule

& & & 	\multicolumn{3}{@{}p{1cm}@{\hskip0.225in}}{\multirow{2}{*}{\textbf{\shortstack{Character\\-istic Sets}}}} 
       & \multicolumn{3}{@{}p{1cm}@{\hskip0.125in}}{\multirow{2}{*}{\textbf{\shortstack{Weak Prop-\\erty Clique}}}}
       & \multicolumn{3}{@{}c@{\hskip0.225in}}{\textbf{SemSets}} 
       & \multicolumn{3}{@{}c@{\hskip0.225in}}{\textbf{SchemEX}} 
       & \multicolumn{3}{@{}c@{\hskip0.225in}}{\textbf{TermPicker}} 
       & \multicolumn{3}{@{}p{1cm}@{\hskip0.125in}}{\multirow{2}{*}{\textbf{\shortstack{SchemEX\\+U+oI}}}} 
       & \multicolumn{3}{@{}p{1cm}@{\hskip0.125in}}{\multirow{2}{*}{\textbf{\shortstack{SchemEX\\+U+pI}}}}
 \\
       
       \\
       
        & \multirow{2}{*}{\textbf{$\bm{k}\ $}} & \multirow{2}{*}{\textbf{Q}}  & \multirow{2}{*}
{\rotatebox[origin=c]{90}{1k}}&\multirow{2}{*}{\rotatebox[origin=c]{90}{100k}}&\multirow{2}{*}{\rotatebox[origin=]{90}{200k}} & \multirow{2}{*}{\rotatebox[origin=c]{90}{1k}}&\multirow{2}{*}{\rotatebox[origin=c]{90}{100k}}&\multirow{2}{*}{\rotatebox[origin=c]{90}{200k}} & \multirow{2}{*}{\rotatebox[origin=c]{90}{1k}}&\multirow{2}{*}{\rotatebox[origin=c]{90}{100k}}&\multirow{2}{*}{\rotatebox[origin=c]{90}{200k}}&\multirow{2}{*}{\rotatebox[origin=c]{90}{1k}}&\multirow{2}{*}{\rotatebox[origin=c]{90}{100k}}&\multirow{2}{*}{\rotatebox[origin=c]{90}{200k}}&\multirow{2}{*}{\rotatebox[origin=c]{90}{1k}}&\multirow{2}{*}{\rotatebox[origin=c]{90}{100k}}&\multirow{2}{*}{\rotatebox[origin=c]{90}{200k}}&\multirow{2}{*}
{\rotatebox[origin=c]{90}{1k}}&\multirow{2}{*}{\rotatebox[origin=c]{90}{100k}}&\multirow{2}{*}{\rotatebox[origin=c]{90}{200k}}&\multirow{2}{*}
{\rotatebox[origin=c]{90}{1k}}&\multirow{2}{*}{\rotatebox[origin=c]{90}{100k}}&\multirow{2}{*}{\rotatebox[origin=c]{90}{200k}}	\\ 
\\

\midrule
\parbox[t]{4mm}{\multirow{6}{*}{\rotatebox[origin=c]{90}{\textbf{TimBL-11M}}}}
& 0 & SQ& $na$&$na$&$na$ & $na$&$na$&$na$ & $.94$&$.97$&$.98$ & $.94$&$.97$&$.98$ & $.94$&$.97$&$.98$ & $.85$&$.91$&$.93$ & $.92$&$.97$&$.98$ \\

\arrayrulecolor{black!30}\cmidrule(r{1pt}){2-24}

& 1 & SQ& $na$&$na$&$na$ & $na$&$na$&$na$ & $.73$&$.76$&$.76$ & $.73$&$.75$&$.76$ & $.22$&$.25$&$.29$ & $.65$&$.70$&$.71$ & $.80$&$.80$&$.85$\\
& 1 & CQ& $.60$&$.77$&$.78$ & $.76$&$.76$&$.76$ & $.39$&$.44$&$.44$ & $.39$&$.44$&$.46$ & $.14$&$.23$&$.29$ & $.33$&$.42$&$.45$ & $.49$&$.50$&$.65$ \\

\arrayrulecolor{black!30}\cmidrule(r{1pt}){2-24}

& 2 & SQ& $na$&$na$&$na$ & $na$&$na$&$na$ & $.71$&$.78$&$.78$ & $.59$&$.77$&$.78$ & $.19$&$.21$&$.24$ & $.41$&$.69$&$.70$ & $.67$&$.73$	&$.74$\\
& 2 & CQ& $.04$&$.18$&$.21$ & $.42$&$.44$&$.45$ & $.26$&$.32$&$.33$ & $.11$&$.24$&$.26$ & $.04$&$.05$&$.07$ & $.06$&$.18$&$.19$ &$.16$&$.22$&$.23$	\\

\arrayrulecolor{black!100}\midrule

\parbox[t]{4mm}{\multirow{6}{*}{\rotatebox[origin=c]{90}{\textbf{DyLDO-127M}}}} 
& 0 & SQ& $na$&$na$&$na$ & $na$&$na$&$na$ & $.56$&$.57$&$.58$ & $.56$&$.57$& $.58$ & $.56$&$.57$&$.58$ & $.44$&$.47$&$.47$ &$.57$&$.59$&$.60$\\

\arrayrulecolor{black!30}\cmidrule(r{1pt}){2-24}

& 1 & SQ& $na$&$na$&$na$ & $na$&$na$&$na$ & $.49$&$.50$&$.50$ & $.49$&$.49$&$.50$ & $.36$&$.37$&$.37$ & $.41$&$.43$&$.43$  &$.88$&$.89$&$.91$	\\
& 1 & CQ&  $.68$&$.71$&$.72$ & $.89$&$.92$&$.89$ & $.31$&$.32$&$.33$ & $.16$&$.17$&$.18$ & $.12$&$.14$&$.15$ & $.13$&$.13$&$.14$ &$.39$&$.41$&$.43$	\\

\arrayrulecolor{black!30}\cmidrule(r{1pt}){2-24}

& 2 & SQ& $na$&$na$&$na$ & $na$&$na$&$na$ & $.49$&$.50$&$.50$ & $.49$&$.49$&$.50$ & $.35$&$.36$&$.36$ & $.41$&$.43$&$.43$ &$.57$&$.59$&$.60$	\\
& 2 & CQ& $.03$&$.13$&$.15$ & $.51$&$.58$&$.58$ & $.23$&$.25$&$.25$ & $.08$&$.09$&$.09$ & $.04$&$.04$&$.04$ & $.06$&$.07$&$.08$ &$.07$&$.08$&$.09$	\\

\arrayrulecolor{black!100}\bottomrule

\end{tabularx}
\label{tab:results}
%}
}
\end{table}
\end{landscape}

\end{document}